# Unusual Isotope Effect on Thermal Transport of Single Layer Molybdenum Disulphide (MoS$_2$)

Xufei Wu[1], Nuo Yang[2,3], Tengfei Luo[1,4,*]

1. Aerospace and Mechanical Engineering, University of Notre Dame

2. School of Energy and Power Engineering, Huazhong University of Science and Technology (HUST), Wuhan 430074, People's Republic of China

3. State Key Laboratory of Coal Combustion, Huazhong University of Science and Technology (HUST), Wuhan 430074, People's Republic of China

4. Center for Sustainable Energy at Notre Dame, University of Notre Dame

\* address correspondence to: tluo@nd.edu

**Abstract:**

Thermal transport in single layer molybdenum disulfide (MoS$_2$) is critical to advancing its applications. In this paper, we use molecular dynamics (MD) simulations with first-principles force constants to study the isotope effect on the thermal transport of single layer MoS$_2$. Through phonon modal analysis, we found that isotopes can strongly scatter phonons with intermediate frequencies, and the scattering behavior can be radically different from that predicted by conventional scattering model based on perturbation theory (Tamura's formula). Such a discrepancy becomes smaller for low isotope concentrations. Natural isotopes can lead to a 30% reduction in thermal conductivity for large size samples. However, for small samples where boundary scattering becomes significant, the isotope effect can be greatly suppressed. It was also found that the Mo isotopes, which contribute more to the phonon eigenvectors in the intermediate frequency range, have stronger impact on thermal conductivity than S isotopes.

Single layer molybdenum disulfide ($MoS_2$) is an attractive two dimensional (2D) material due to its intrinsic bandgap of 1.8 ev and high electron mobility around 200 $cm^2V^{-1}S^{-1}$, which can potentially enable applications in transistors, photovoltaics and valleytronics.[1-5] Due to the extremely thin geometry of single layer $MoS_2$, a small amount of Joule heating can lead to large temperature rises, impairing the device performance and reliability. This makes the understanding of thermal transport in $MoS_2$ of great importance. On the other hand, recent work shows that $MoS_2$ can have very large Seebeck coefficient ranging from $-4\times10^2$ to $-1\times10^5$ μV/K, shedding lights on its promise in thermoelectric applications.[6] In this aspect, the ability to reduce thermal conductivity is valuable since a low thermal conductivity can potentially enhance the energy conversion efficiency of thermoelectrics.

The thermal conductivity of few-layer and single-layer $MoS_2$ has been studied recently. Yan et al.[7] measured the thermal conductivity of suspended single layer $MoS_2$ by Raman spectroscopy and found the value to be around 34.5 W/mK. Using the same method, Sahoo et al.[8] reported that the thermal conductivity of few layer $MoS_2$ was ~52 W/mK. Liu et al.[9] used pump-probe to measure thermal conductivity of bulk $MoS_2$ and found the in-plane value to be 85~110 W/mK.

Wei et al.[10] calculated the thermal conductivity of single layer $MoS_2$ using Klemens' formula[11, 12] and found the value to be 26.2 W/mK at room temperature. Klemens' formula, however, involves significant approximation on the phonon scattering phase space, which can lead to large error in thermal conductivity prediction. Theoretically, 2D materials should have divergent thermal conductivity as a function of length due to the special 2D feature of the phonon dispersion, which prevents some low frequency phonons to be scattered.[13] Gu et al.[14] used iterative solution of phonon Boltzmann transport equations (BTE) with first-principles lattice dynamics and found the thermal conductivity at room temperature to be around 100 W/mK with a sample length of 1 μm.

Cepellotti *et al.*[15] used similar methods and found the thermal conductivity without length effect to be 400 W/mK at room temperature. Both of these studies found that the iterative solution of BTE provides higher thermal conductivity than those predicted using the single mode relaxation time approximation (SMRTA). This is because SMRTA includes both Normal and Umklapp processes as resistive scatterings, but the Normal processes conserve momentum which thus should not lead to thermal resistance. The discrepancy between these methods becomes smaller in smaller samples, where boundary scattering becomes more dominant.[14] Li *et al.*[16] studied the size effect of $MoS_2$ from first-principles calculations using SMRTA and found the thermal conductivity of 1 μm to be 83 W/mK and can be increased by 30% for sample with a size of 10 μm.

Molecular Dynamics (MD) simulations with empirical potentials have also been used to calculate the thermal conductivity of $MoS_2$.[17-19] However, the predicted results range from 5 to 117 W/mK. The lack of convergent data from different MD simulations is likely due to the fact that it is very hard for the empirical potentials to precisely describe both the harmonic and anharmonic properties of the interactions, which are both important for thermal conductivity prediction.

Despite all these efforts in pure $MoS_2$, no study has been performed to investigate the isotope effect. Generally, isotopes can reduce thermal conductivity by phonon scatterings while maintaining the electronic properties,[20] which is attractive for thermoelectric applications. On the other hand, achieving isotope-enriched samples will reduce such phonon scattering and thus improve thermal conductivity, which will be beneficial for thermal management of electronics self-heating.

In this paper, we study the effect of isotope on the thermal conductivity of single layer $MoS_2$ using MD simulations. To ensure the accuracy of the MD potential, we extract the interatomic force constants from first-principles calculations, and used the second and third order force constants in MD simulations to describe the interatomic interactions (see supplementary information (SI)).[21] This method has led to accurate prediction of phonon dispersion and thermal conductivity for different materials.[22-27] Combined with phonon modal spectral energy density (SED) analysis, MD simulations are performed to extract the phonon relaxation times, and the effect of isotope on the relaxation times is studied. The relaxation times due to isotope scattering obtained from the MD simulations are found drastically different from those predicted by the conventionally used Tamura's isotope scattering formula especially at high isotope concentrations.[28] It is found that natural isotopes can greatly reduce the thermal conductivity. By tuning the type and concentration of isotopes artificially, further thermal conductivity reduction can also be achieved.

We first calculate the thermal conductivity of single layer $MoS_2$ without isotopes using NEMD to validate our model. It is well-known that in NEMD simulations, there is a finite size effect which effectively limits the phonon mean free path (MFP) to the distance between the two thermostats which scatter phonons artificially (**Fig. 1a**).[22, 29, 30] Phonons, which have intrinsic MFPs longer than the simulation domain size can transport ballistically, and their contributions to the apparent thermal conductivity are suppressed.[31, 32] Such a classical size effect provides the opportunity to study the thermal conductivity as a function of sample length, which is shown in **Fig. 1b**. There is an obvious increase in thermal conductivity when the sample length increases, in which case longer MFP phonons start to contribute to thermal conductivity. Compared with the reference data from first-principles calculations,[14, 16] our NEMD results show a generally good agreement. Due to limitations on the computational resource, samples longer than 200 nm are not simulated.

To extend the range of length, we performed SED analysis[33, 34] to extract the phonon relaxation times (**Fig. 2a**) and used **Eq. 1** below to calculate the thermal conductivity as a function of length. Details of NEMD simulation and SED analysis are included in SI.[21]

$$\kappa(\omega) = \int c_v(\omega) v(\omega)^2 \tau(\omega) d\omega \tag{1}$$

where $c_v = k_B/V$ is the volumetric heat capacity per mode at the classical limit, $v$ is the phonon group velocity and $\tau$ is the relaxation time. The classical heat capacity is used for a fair comparison between the SED results and the NEMD results since our MD simulations are classical. The effective relaxation time, $\tau$, is calculated as $\tau^{-1} = \tau_{ph-ph}^{-1} + v/L$, where $\tau_{ph-ph}$ is the relaxation time due to intrinsic phonon-phonon interaction, which is obtained from SED, and the second term refers to the boundary scattering effect imposed by the finite sample length, $L$. This enables us to calculate the thermal conductivity as a function of sample length. It is seen from **Fig. 1b** that the thermal conductivity predicted from SED is consistent with our NEMD results and generally agrees with the first-principles calculations[14,16] over a large length range from 5 nm to 10 μm. It is worth mentioning that similar to SMRTA, the SED method assigns a relaxation time for each phonon mode, which should underestimate thermal conductivity compared to that from the NEMD simulations. However, due to the size effect in NEMD (50 – 2000 Å), this error can be overshadowed by the boundary scattering, and thus we see little difference between NEMD and SED results.

Our MD predictions are also comparable to experimental results although they do not have a unified value among themselves. A recent pump-probe measurement of bulk $MoS_2$ yielded the in-plane thermal conductivity of 85-110W/mK.[9] Although not specifically characterized in Ref. 9, natural $MoS_2$ grain sizes are usually ~1-2 μm.[35, 36] Corresponding to this length, our predicted thermal conductivity using the SED is ~140 W/mK. Considering the interlayer phonon scattering in the bulk samples and the isotope scattering in experimental samples, we regard this as a good agreement between our MD prediction and this pump-probe experimental datum.

With the confidence in our model, we then investigate natural isotope effect on the thermal conductivity of single layer $MoS_2$. Tamura's formula[28] (**Eq. 2**) derived from perturbation theory is usually employed to describe isotope scattering, and it has been used to study isotope effect in 2D materials.[15, 37]

$$\tau_{iso}^{-1}(\lambda) = \frac{\pi \omega_\lambda^2}{2N} \sum_{\lambda'} \delta(\omega_\lambda - \omega_{\lambda'}) \sum_\sigma g(\sigma) \left| e^*(\sigma|\lambda') \cdot e(\sigma|\lambda) \right|^2 \qquad (2)$$

where $\omega_\lambda$ is the angular frequency of mode $\lambda$, $N$ is the number of $k$-point sampled in the reciprocal space, $e(\sigma|\lambda)$ is the eigenvector of mode $\lambda$ at atom $\sigma$. The mass variance $g$ is given by $g(\sigma) = \sum_s f_s (1 - M_s / \bar{M})^2$, where $f_s$ is the atomic fraction of the $s$ th isotope of atom $\sigma$, $M_s$ and $\bar{M}$ are the masses of the $s$ th isotope atom and the normal atoms, respectively. We take the relaxation times from the SED analysis on pure $MoS_2$ (blue circles, **Fig. 2a**) and impose the isotope scattering (**Eq. 2**) using Matthiessen's rule, $\tau^{-1} = \tau_{ph-ph}^{-1} + \tau_{iso}^{-1}$, by considering the natural isotope composition. It is seen from **Fig. 2a** (red circles) that the effect of isotope scattering increases dramatically when the frequency increases. Based on these relaxation times, we calculated the cumulative thermal conductivity as a function of frequency according to **Eq. 1** and found that the thermal conductivity can be dramatically reduced (**Fig. 2b**).

It is worth noting that **Eq. 2** assumes that the concentration of the isotopes approaches the dilute limit. However, it is known that the isotope phonon scattering also depends on the density and the clustering of isotopes,[37] and thus **Eq. 2** could lead to inaccuracy. We then simulate single layer

MoS$_2$ with natural isotope composition in MD, and use SED to extract the phonon relaxation times (pink circles, **Fig. 2a**). It is found that the relaxation times estimated from Tamura's formula are uniformly smaller than those extracted from SED, which leads to a difference of 70 W/mK in total thermal conductivity. This suggests that the use of the Tamura formula can lead to inaccuracy for predicting thermal conductivity of MoS$_2$ with natural isotope compositions. It is worth mentioning that this inaccuracy can be from either the Tamura formula itself or the Matthiessen's rule, but we cannot separate their effects.

As can be seen, the difference in relaxation time from the SED and Tamura's formula becomes obvious when frequency is larger than 2 THz, and the difference is generally larger at higher frequencies. Phonons with frequencies between 2 to 7 THz are the main contributors to the total thermal conductivity in (**Fig. 2b**). Thus, the relaxation time difference in this mid-frequency range is responsible for the large thermal conductivity difference between the two different methods. For phonons with higher frequency, the inaccuracy of the Tamura formula does not contribute much to the error in thermal conductivity (**Fig. 2b**). The reason is that the phonon group velocities of optical phonons are much smaller compared to the low frequency acoustic modes (**Fig. 2c**). Considering that the modal contribution to thermal conductivity scales with $v^2$ (**Eq. 1**), the effect of the relaxation time difference is largely suppressed. This is also the reason of the very flat profiles after 7 THz in the cumulative thermal conductivities (**Fig. 2b**).

To further evaluate the applicability of the Tamura formula, we calculated the relaxation times and thermal conductivities for different Mo isotope concentrations. As shown in **Fig. S6** in SI,[21] the accuracy of the Tamura formula improves as the isotope concentration becomes lower, but the discrepancy is already noticeable when the concentration is 5% (**Fig. S6**). The application of Tamura's formula can be reasonable for low isotope concentrations (dilute limit), but at higher concentrations large errors can be expected.

The above discussion on isotope scattering has not yet involved size effect which can compete with the isotope effect. In **Fig. 3a** and **3b**, we plot relaxation times for two different sample lengths and illustrate this competing effect. Comparing these two figures, it is clear that for the short sample ($L = 200$ nm), length effect (boundary scattering) dominates the phonon scattering for acoustic phonons, and the isotope scattering effect is significantly suppressed. The only obvious impact from isotope is at the high frequency (optical phonons) region. However, since these phonons do not contribute much, the thermal conductivity for the $L = 200$ nm sample only show a ~5% reduction when the isotopes are introduced (**Fig. 3c**). We have also calculated the thermal conductivity of the isotopic $MoS_2$ using NEMD directly for $L = 206$ nm – the largest length we can simulate – and found the thermal conductivity is reduced to 38 W/mK from 41 W/mK (a ~7% reduction), which agree reasonably with the SED prediction. However, for the longer sample ($L = 2\mu m$), the isotope scattering becomes more significant especially for the mid-frequency region (boxed in **Fig. 3b**). As discussed before, these modes contribute significantly to the thermal conductivity and thus the natural isotope scattering can lead to a large reduction in thermal conductivity by ~ 30% (**Fig. 3c**).

With both isotope effect and length effect considered in **Fig. 3c**, the thermal conductivity of natural $MoS_2$ for samples with grain sizes of 1-2μm here is expected to be 95-120 W/mK. This range matches better with the pump-probe measurement by Liu *et al.*[9] on the in-plane thermal conductivity of bulk $MoS_2$. More importantly, our analyses show that for normal

MoS$_2$ samples which have grain sizes on the order of microns, the isotope scattering effect can be significant.

Besides isotopes with natural compositions, we also studied artificial isotope concentrations which can be achieved in synthesis methods like chemical vapor deposition and molecular beam epitaxy. We choose the largest differences in mass of the different isotopes (*Mo$^{100}$ and Mo$^{92}$; S$^{36}$ and S$^{32}$*) and mix them with different ratios. The results are shown in **Fig. 4a**. Even though the large uncertainties prevent very quantitative analysis of the data, the overall trends generally agree with those observed in alloy effect on thermal conductivity:[38] when the mixing ratio changes, thermal conductivity drops first, followed by a flat region, and then increases.

It seems that the Mo isotopes have a larger impact on the thermal conductivity than the S isotopes. To explain this, we extract phonon relaxation times for systems with 50% Mo and 50% S isotopes, respectively, using both SED analysis and the isotope scattering rates from Tamura's formula (**Fig. 4b** and **4c**). Despite the difference between methods, both sets of relaxation times show the same trend: S isotopes can more easily influence high frequency modes (*i.e.*, optical phonons from 8 to 14 THz) which do not contribute much to thermal conductivity. In contrast, Mo isotopes play a more significant role in the frequency range of 2-7 THz. Their impact to phonons in this frequency range can lead to larger effect in thermal conductivity since these phonons dominate thermal transport. The cause of these observations can be traced back to the phonon eigenvectors included in the Tamura's formula (**Eq. 2**): Mo, as the heavier atom, contributes to the eigenvectors of low frequency phonon modes more than the S atom. The largest thermal conductivity reduction when Mo isotopes are introduced is ~ 20%, which is larger than the 7-10% achieved using natural isotopes for the same sample length. It is worth noting that when the same study is performed for samples with much longer lengths, it is expected that the thermal conductivity reduction would be further larger.

In this paper, we studied the isotope effect in the thermal conductivity of single layer $MoS_2$. We found that the natural isotopes can reduce the thermal conductivity by as much as ~30%. Such reduction is mainly due to the isotope scattering of phonons with mid-range frequencies. It is also found that the conventionally used Tamura's scattering formula can lead to large error in relaxation time prediction when isotope concentration is large. The error becomes smaller when isotope concentration is very small (1%). Isotope effect is more pronounced when the sample length is large where the boundary scattering effect is small. By tuning the isotope compositions artificially, larger thermal conductivity reduction can potentially be achieved, and it is found that Mo isotope is more effective in thermal conductivity reduction. These results provide important guidance on the synthesis of single layer $MoS_2$ with desirable thermal conductivity for applications like thermoelectrics and electronics.

**ACKNOWLEDGMENT:**

This research was supported in part by NSF (2DARE 1433490), the Notre Dame Center for Research Computing, and NSF through XSEDE computing resources provided by SDSC Trestles and TACC Stampede under grant number TG-CTS100078. We also thank Dr. Keivan Esfarjani for useful discussions.

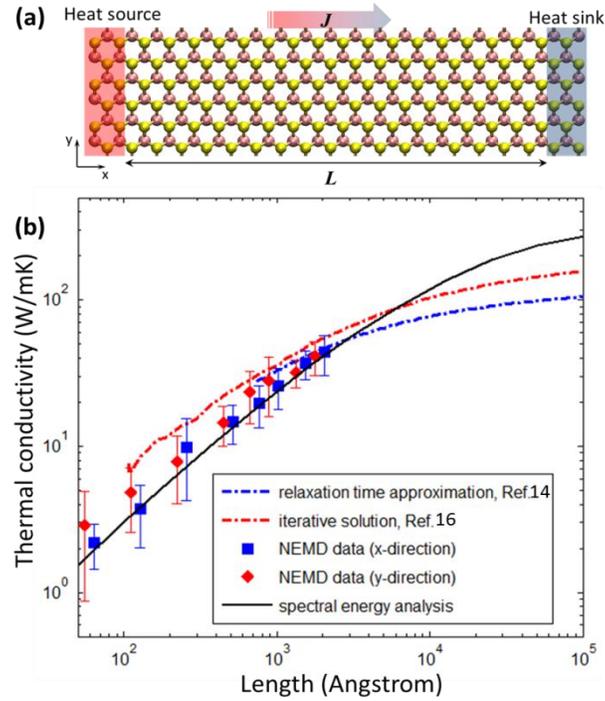

*Figure 1.* *(a) Schematic of NEMD simulation for MoS$_2$ thermal conductivity calculations. (b) Thermal conductivity as a function of length at quantum corrected room temperature. The blue and red dashed lines are from references 14 and 16, respectively. Sample size in the x-direction is from 64.29 to 2057.14 Å and that in the y-direction is from 55.67 to 1781.53 Å.*

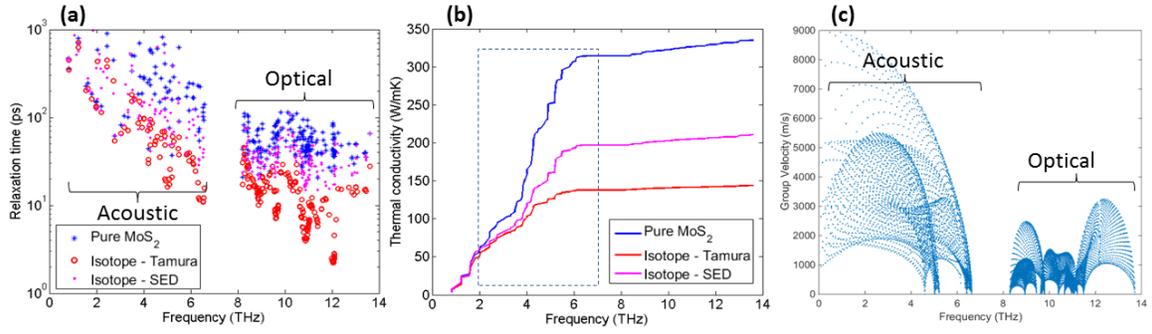

*Figure 2.* (a) Phonon relaxation times of single layer $MoS_2$ with and without natural isotopes; (b) cumulative thermal conductivity as a function of frequency; (c) phonon group velocity as a function of frequency. Sample size in MD simulations is 32.14x82.51 $Å^2$.

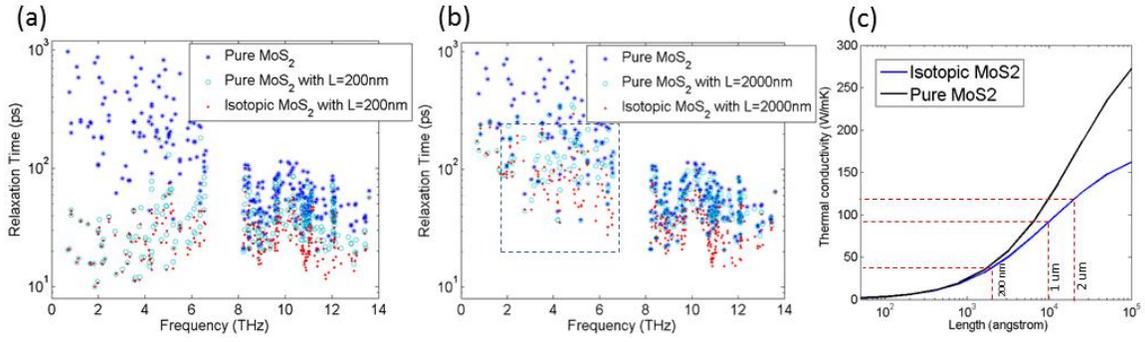

*Figure 3.* Relaxation times under the influences of isotope and boundary scattering effects with (a) L=200 nm and (b) L= 2000 nm; (c) Thermal conductivity as a function of length with and without natural isotope scattering effects.

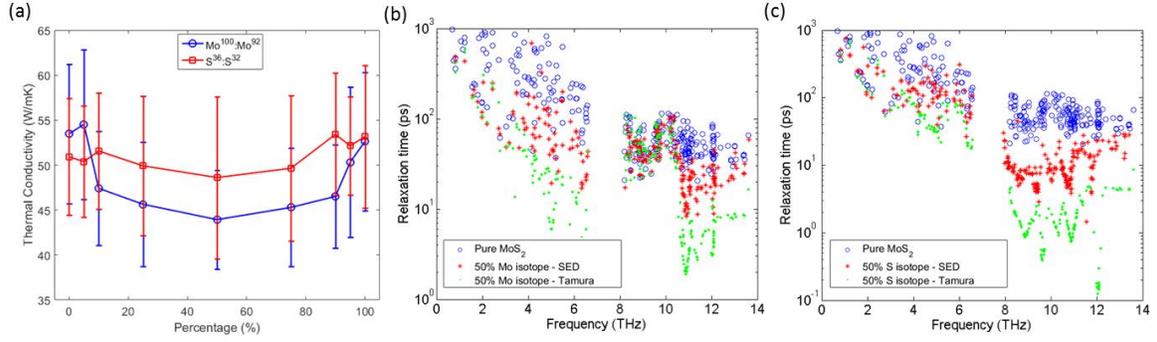

*Figure 4.* *(a) Thermal conductivity as a function of isotope composition for Mo isotopes ($Mo^{100}$ and $Mo^{92}$) and S isotopes ($S^{36}$ and $S^{32}$). Sample size of NEMD simulation is 2057.14x55.67 $Å^2$. The heat flow direction is in the x-direction. (b) and (c) Relaxation time as a function of frequency for Mo isotopes (50% $Mo^{100}$ and 50% $Mo^{92}$) and S isotopes (50% $S^{36}$ and 50% $S^{32}$), respectively. Sample size of the MD simulation is 32.14x82.51 $Å^2$.*

# Supplementary information for

# Unusual Isotope Effect on Thermal Transport of Single Layer Molybdenum Disulphide (MoS$_2$)

Xufei Wu[1], Nuo Yang[2,3], Tengfei Luo[1,4,*]

1. Aerospace and Mechanical Engineering, University of Notre Dame

2. School of Energy and Power Engineering, Huazhong University of Science and Technology (HUST), Wuhan 430074, People's Republic of China

3. State Key Laboratory of Coal Combustion, Huazhong University of Science and Technology (HUST), Wuhan 430074, People's Republic of China

4. Center for Sustainable Energy at Notre Dame, University of Notre Dame

\* address correspondence to: tluo@nd.edu

## 1. Interatomic interaction from first-principles force constants

We use MD with the interatomic interactions expressed by force constants. The interatomic potential, $V$, and force, $F$, can be expressed by the displacements, $u$, of atoms from their equilibrium position using Taylor series expansion. Here, we show the expansion up the third order.

$$V = V_0 + \sum_{i_\alpha} \Omega_{i_\alpha} u_{i_\alpha} + \sum_{i_\alpha j_\beta} \Psi_{i_\alpha j_\beta} u_{i_\alpha} u_{j_\beta} + \sum_{i_\alpha j_\beta k_\gamma} \Phi_{i_\alpha j_\beta k_\gamma} u_{i_\alpha} u_{j_\beta} u_{k_\gamma} \qquad (S1)$$

$$F_{i_\alpha} = -\frac{\partial V}{\partial u_{i_\alpha}} = -\Omega_{i_\alpha} - \sum_{j_\beta} \Psi_{i_\alpha j_\beta} u_{j_\beta} - \frac{1}{2} \sum_{j_\beta k_\gamma} \Phi_{i_\alpha j_\beta k_\gamma} u_{j_\beta} u_{k_\gamma} \qquad (S2)$$

where $i, j, k$ denote the indices of atoms, and $\alpha, \beta, \gamma$ denote the direction (*i.e.*, $x, y, z$). $\Omega, \Psi, \Phi$ represent the 1st, 2nd, 3rd order force constants, respectively. Higher order force constants can be included but they are known to have limited impact on thermal conductivity at intermediate temperatures.[1,2] It is worth mentioning that in some material systems, truncating the potential at the third order term will lead to instability of the system at room or high temperatures.[3] For MoS$_2$, the system is found to be stable at temperatures below 1000 K with the second and the third order terms included.

The force constants are extracted from first-principles DFT calculations. The DFT calculations are performed using the planewave based Quantum-Espresso package.[4] Ultrasoft pseudopotentials with Perdew-Wang 91 Generalized Gradient Approximation[5] are used for both Mo and S atoms. A planewave cut-off of 50 Rydberg and a Monkhorst-Pack[6] mesh of 2×2×1 in the *k*-space are chosen based on the convergence test of the lattice energy. The force constants are extracted from the force-displacement data from the DFT calculations. Up to fourth nearest neighbors interactions are included for the second order (harmonic) force constant, and the nearest neighbors are included for the cubic (anharmonic) terms. A 5×5 supercell containing 75 atoms is used for the calculations (Fig. S1). The details of the method to extract the force constants can be found in Ref.[1]

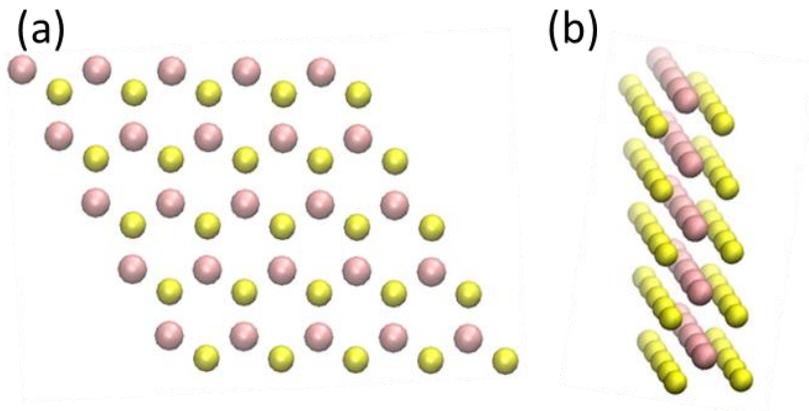

(a) (b)

**Figure S1. The (a) top view and (b) side view of the single layer MoS$_2$ supercell used in DFT calculations for force constants extraction.**

It is notable that not all of the force constants are independent of each other. The translation invariance and space group symmetries must be considered for the specific lattice structure studied. Failure to consider the symmetry relations of force constants can result in extra phonon scattering channels and instability in MD simulations. The symmetries we applied include translational invariance of system (Eq. (S3)) and invariance under symmetry operations (Eq. (S4)):

$$\sum_i \Omega_{i_\alpha} = 0; \quad \sum_j \Psi_{i_\alpha j_\beta} = 0; \quad \sum_k \Phi_{i_\alpha j_\beta k_\gamma} = 0 \qquad (S3)$$

$$\Omega_{(Si)_\alpha} = \sum_{\alpha'} \Omega_{i_{\alpha'}} W_{\alpha,\alpha'}; \quad \Psi_{(Si)_\alpha (Sj)_\beta} = \sum_{\alpha'\beta'} \Psi_{i_{\alpha'} j_{\beta'}} W_{\alpha,\alpha'} W_{\beta,\beta'}; \quad \Phi_{(Si)_\alpha (Sj)_\beta (Sk)_\gamma} = \sum_{\alpha'\beta'\gamma'} \Phi_{i_{\alpha'} j_{\beta'} k_{\gamma'}} W_{\alpha,\alpha'} W_{\beta,\beta'} W_{\gamma,\gamma'}.$$
(S4)

Here $W$ is the 3×3 matrix of symmetry operation. $S_i$ is the index of a new atom by symmetry operation of atom $i$. All the force constants extracted from first-principles calculations are listed in Tables S1 & S2.

**Table S1. Second order Force constants of MoS$_2$**

| Index of atoms | | Directions | | Second order force constants (ev/ Å$^2$) |
|---|---|---|---|---|
| 38 | 38 | z | z | 25.8358 |
| 38 | 38 | x | x | 24.2047 |
| 37 | 37 | z | z | 14.8399 |
| 37 | 37 | x | x | 11.1212 |
| 37 | 35 | y | y | -5.22803 |
| 37 | 35 | z | z | -4.66553 |
| 37 | 38 | x | x | -4.31175 |
| 37 | 35 | y | z | 3.59455 |
| 37 | 38 | x | z | 3.11297 |
| 37 | 38 | y | y | -2.47917 |

| 37 | 36 | z | z | -1.82271 |
|---|---|---|---|---|
| 37 | 38 | y | z | -1.79727 |
| 37 | 38 | x | y | 1.58706 |
| 37 | 35 | x | x | -1.56288 |
| 38 | 23 | x | x | -1.19522 |
| 38 | 20 | y | y | -0.96316 |
| 37 | 22 | x | x | -0.66598 |
| 38 | 20 | z | z | 0.618 |
| 38 | 20 | x | x | -0.49903 |
| 37 | 19 | y | y | -0.47096 |
| 38 | 20 | x | y | -0.40195 |
| 37 | 56 | x | x | 0.400118 |
| 37 | 22 | z | z | 0.350834 |
| 37 | 19 | x | y | -0.33779 |
| 37 | 36 | y | y | 0.312301 |
| 38 | 23 | y | y | -0.26696 |
| 38 | 54 | z | z | -0.25845 |
| 37 | 20 | y | y | 0.246279 |
| 37 | 56 | y | y | -0.21524 |
| 37 | 21 | y | y | 0.190123 |
| 37 | 21 | x | z | 0.158639 |
| 37 | 21 | x | x | -0.1464 |
| 37 | 18 | x | y | -0.14572 |
| 37 | 18 | y | z | 0.137386 |
| 37 | 56 | y | z | -0.12634 |
| 37 | 22 | y | y | 0.114102 |
| 37 | 20 | x | z | 0.109417 |
| 37 | 18 | x | x | 0.105993 |
| 37 | 19 | x | x | -0.08092 |
| 37 | 18 | x | z | 0.07932 |
| 37 | 50 | y | z | 0.063172 |
| 38 | 24 | x | x | -0.0614 |
| 37 | 18 | z | z | -0.05838 |
| 37 | 22 | y | z | 0.022347 |
| 37 | 19 | x | z | 0.019353 |
| 37 | 19 | y | z | -0.01117 |

**Table S2. Third order Force constants of MoS$_2$**

| Index of atoms | Directions | Third order force constants (ev/ Å$^3$) |
|---|---|---|

| 37 | 37 | 37 | z | z | z | 23.2116 |
|---|---|---|---|---|---|---|
| 37 | 37 | 37 | y | y | z | 19.9468 |
| 38 | 38 | 38 | x | x | y | -13.0627 |
| 38 | 39 | 39 | y | y | z | 12.3695 |
| 38 | 21 | 21 | x | x | z | 9.50986 |
| 37 | 35 | 35 | y | z | z | -9.269 |
| 37 | 35 | 35 | y | y | y | -8.56237 |
| 37 | 53 | 53 | x | z | z | 8.0278 |
| 37 | 53 | 53 | z | z | z | 7.73769 |
| 37 | 37 | 37 | x | x | y | 6.53126 |
| 37 | 38 | 38 | x | x | x | -5.53306 |
| 38 | 21 | 21 | x | y | z | 4.95375 |
| 37 | 53 | 53 | y | z | z | 4.63485 |
| 38 | 36 | 36 | y | y | z | 3.79053 |
| 38 | 21 | 21 | x | x | y | -3.2421 |
| 37 | 38 | 38 | x | y | y | -1.84389 |
| 38 | 36 | 36 | y | y | y | -1.01683 |
| 37 | 35 | 35 | x | x | z | 0.929844 |

Table S3. The corresponding position of atoms with index used in Table 1 and 2.

| Atom index | Position (x,y,z) (unit: Å) | | |
|---|---|---|---|
| 18 | 0 | -3.71153 | 3.16044 |
| 19 | 0 | -3.71153 | 0 |
| 20 | -1.60714 | -2.78365 | 1.58022 |
| 21 | -1.60714 | -0.92788 | 3.16044 |
| 22 | -1.60714 | -0.92788 | 0 |
| 23 | -3.21428 | 0 | 1.58022 |
| 24 | -3.21428 | 1.85576 | 3.16044 |
| 35 | 1.60714 | -2.78365 | 1.58022 |
| 36 | 1.60714 | -0.92788 | 3.16044 |
| 37 | 1.60714 | -0.92788 | 0 |
| 38 | 0 | 0 | 1.58022 |
| 39 | 0 | 1.85576 | 3.16044 |
| 50 | 4.82142 | -2.78365 | 1.58022 |
| 53 | 3.21428 | 0 | 1.58022 |
| 54 | 3.21428 | 1.85576 | 3.16044 |
| 56 | 1.60714 | 2.78365 | 1.58022 |

We validate the accuracy of the harmonic force constants by comparing the calculated phonon dispersion relation to reference data (Fig. S2). It is seen that our calculated phonon dispersion agrees with the reference data favorably, indicating the accuracy of the harmonic force constants. The accuracy of the cubic force constants is indicated by the calculated thermal conductivity in the main text which agrees well with other first-principles lattice dynamics calculations.

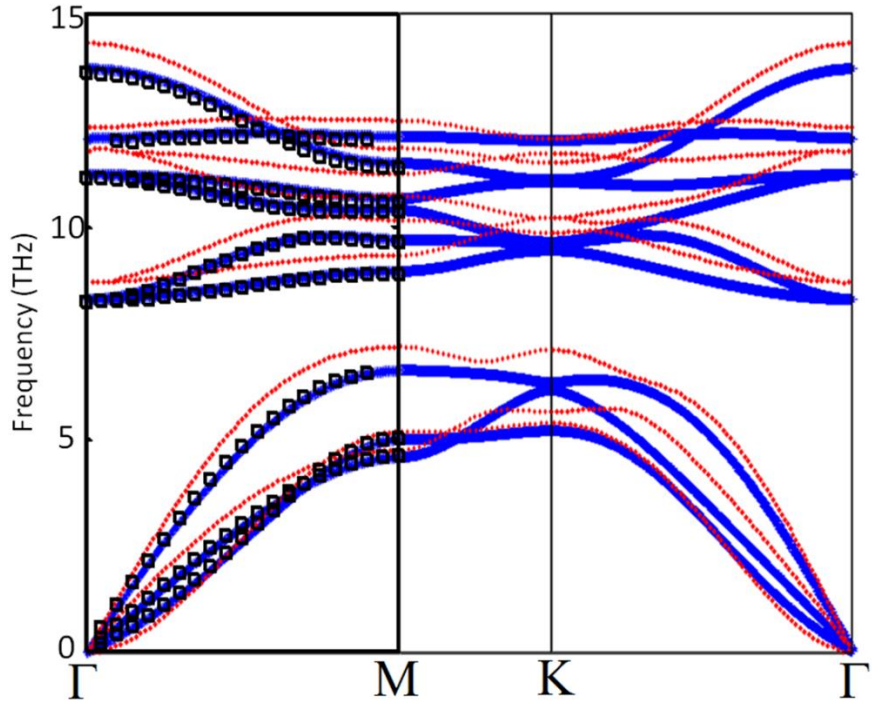

**Figure S2. Phonon dispersion relation of single layer $MoS_2$. Reference data[7] are plotted in red dashed line. The black squares are frequencies extracted from the SED analysis.**

**2. Molecular dynamics (MD) simulation details**

To calculate the thermal conductivity of $MoS_2$, we use non-equilibrium MD (NEMD), in which we apply a temperature gradient across the sample in the x-direction by setting each end of the sample at different temperatures using Langevin thermostats (**Fig. 1a**). At the steady state, the heat flux ($J$) can be extracted by monitoring the rate of energy input and output at the thermostats. Thermal conductivity is then calculated according to Fourier's law, $k = -J/\nabla T$, with the temperature gradient, $\nabla T$, being the slope of the linear portion of the steady state temperature. The simulations are carried out at 320 K, which corresponds to room temperature after quantum correction.[8] The temperatures of the thermostats are set to 330 K and 310 K, respectively. We apply periodic boundary conditions along the width direction and fixed boundary conditions in the heat transfer direction. The sample width is fixed as 55.67 Å, while the length varies from 64.29 to 2057.14 Å. Thermal conductivity in the y-direction is also calculated due to expected anisotropy. The width in these simulations is fixed as 64.29 Å, and the length varies from 55.67 to 1781.53 Å. Our convergence test shows that the widths chosen are large enough to give converged thermal conductivity values. In all simulations, the cutoff distances of harmonic and anharmonic interactions are up to the 4th nearest and the nearest neighbor shells, respectively. An equilibration run in the NVT ensemble (constant number of atoms, volume and temperature) is carried out for 50 ps, followed by a preproduction run in the NVE ensemble (constant number of atoms, volume and energy) for 200 ps, after which data are collected for 200 ps. For each simulation, 20 independent runs with different initial conditions are performed, and the thermal conductivities are averaged. The error bar is calculated as the standard deviation.

## 3. Quantum correction of temperature

To calculate the thermal conductivity, we take the quantum effects into account, which is important at temperatures below the Debye temperature and is usually not considered in the classical MD simulations. The quantum temperature of the system can be related to MD temperature by:[8]

$$3Nk_bT_{MD} = \int_0^{\omega_{max}} D_{tot}(\omega)\left[\frac{1}{e^{\hbar\omega/k_bT}-1}+\frac{1}{2}\right]\hbar\omega d\omega \qquad (S5)$$

where $D_{tot}(\omega)$ is the phonon density of states summed overall all acoustic branches. Here the Debye model is applied and Eq. (S5) is converted into:

$$T_{MD} = \frac{1}{3k_b}\int_0^{\upsilon_D} b_{tot}(\omega)\left[\frac{1}{e^{h\upsilon/k_bT}-1}+\frac{1}{2}\right]h\upsilon d\upsilon \qquad (S6)$$

where the total density of states is summed over all three acoustic branches:

$$b_{tot} = b_{LA}+b_{TA}+b_{ZA} = \frac{4\pi\upsilon^2}{N/V}\left(\frac{1}{c_{LA}^3}+\frac{1}{c_{TA}^3}+\frac{1}{c_{ZA}^3}\right) = \frac{4\pi\upsilon^2}{N/V}\frac{3}{c_{av}^3} \qquad (S7)$$

and the Debye frequency $\upsilon_D$ can be evaluated as:

$$\upsilon_D = c_{av}\left(\frac{3N}{4\pi V}\right)^{1/3} \qquad (S8)$$

Here the velocities of the individual branches are calculated to be 5760.97, 3410.85 and 2425.93 m/s, resulting in an averaged velocity $c_{av} = 3102.34 m/s$. The relation of MD temperature and quantum temperature is plotted in Figure S3. In our MD simulations, a MD temperature of 320 K corresponds to the quantum temperature of room temperature.

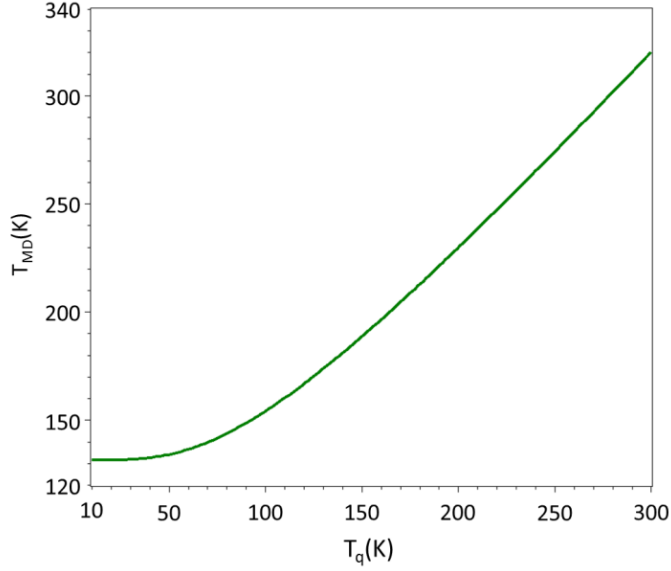

**Figure S3. MD temperature as a function of quantum temperature.**

### 4. Phonon relaxation time from spectral energy density analysis

In order to calculate phonon relaxation times, equilibrium MD simulations are performed. The simulations run at 320 K. We first calculate the spectral energy density function, which is the average kinetic energy per unit cell as a function of wave vector. By recording the velocities of atoms as a function of time, the spectral energy density can be calculated as:[9]

$$\Phi(\kappa,\omega) = \frac{1}{4\pi\tau_0 N_T} \sum_{\alpha \in \{x,y,z\}} \sum_{b} m_b \left| \int_0^{\tau_0} \sum_{n_{x,y,z}} \dot{u}_\alpha\left(n_{x,y,z}, b; t\right) \times \exp\left[i\kappa \cdot r\left(n_{x,y,z}\right) - i\omega t\right] dt \right|^2 \quad (S9)$$

where $\tau_0$ is the integration time constant, $N_T$ is the total number of unit cells in the system, $\alpha$ represents the direction of velocity, $b$ represents the type of atoms in the unit cell, $n_{x,y,z}$ represents different unit cells and $r(n_{x,y,z})$ represents the vector of displacement between unit cell $n_{x,y,z}$ and the basis unit cell. We extract the relaxation time of phonon modes by fitting the spectral energy density to a linear superposition of Lorentzian functions, whose peaks and linewidths correspond to the fully anharmonic phonon frequency and phonon relaxation time:

$$\Phi(\omega) = \sum_i \frac{I(\omega_i)}{\left[2\tau_i(\omega-\omega_i)\right]^2 + 1} \tag{S10}$$

Here $I(\omega_i)$ is the peak value and $\tau_i$ is the relaxation time of a phonon mode i. An example of the fitted spectral energy density is shown in Fig. S4.

It should be noticed that phonon modes cannot be fitted at all K-points. The condition $\exp(iK \cdot R) = 0$ has to be satisfied, where R is the periodicity of boundary conditions in MD domain. Here we apply a 10x30 sample size with periodic boundary conditions in both x and y directions. And a set of mesh points satisfying the boundary condition are picked as shown in figure S5. The irreducible first Brillouin zone on a 41 point mesh and the data are then mapped into the full first Brillouin zone.

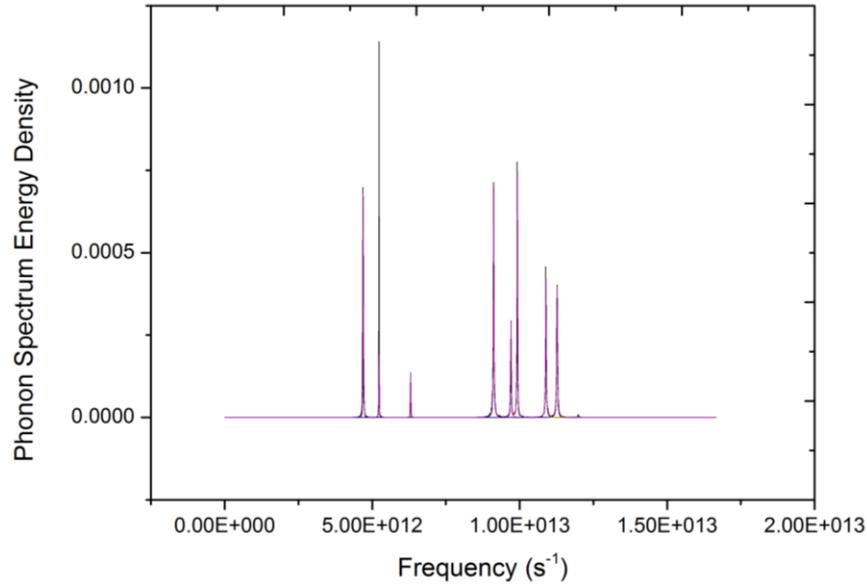

**Figure S4. Phonon spectral energy density and the fitted peaks.**

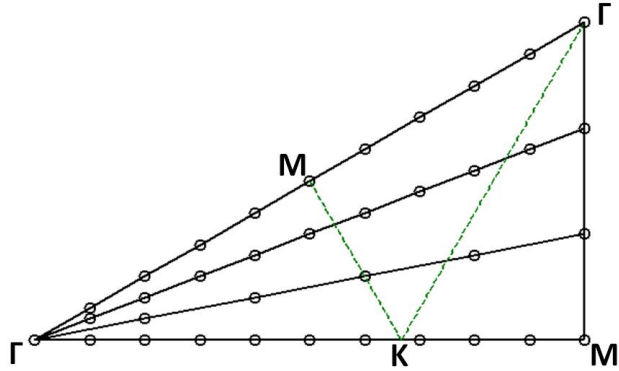

**Figure S5. Meshing points in reciprocal K space**

## 5. Comparison of relaxation time from SED analysis and perturbation theory under different isotope concentration condition

To further evaluate the applicability of the Tamura formula, we have also calculated the relaxation times and thermal conductivity for different Mo isotope concentrations. As shown in Fig. S6 in the supplemental file, the accuracy of the Tamura formula improves as the isotope concentration becomes lower, but the discrepancy is already noticeable when the concentration is 5% (Fig. S6). It can be concluded that the application of Tamura formula is reasonable for low isotope concentrations (dilute limit), but at higher concentrations large error can be expected.

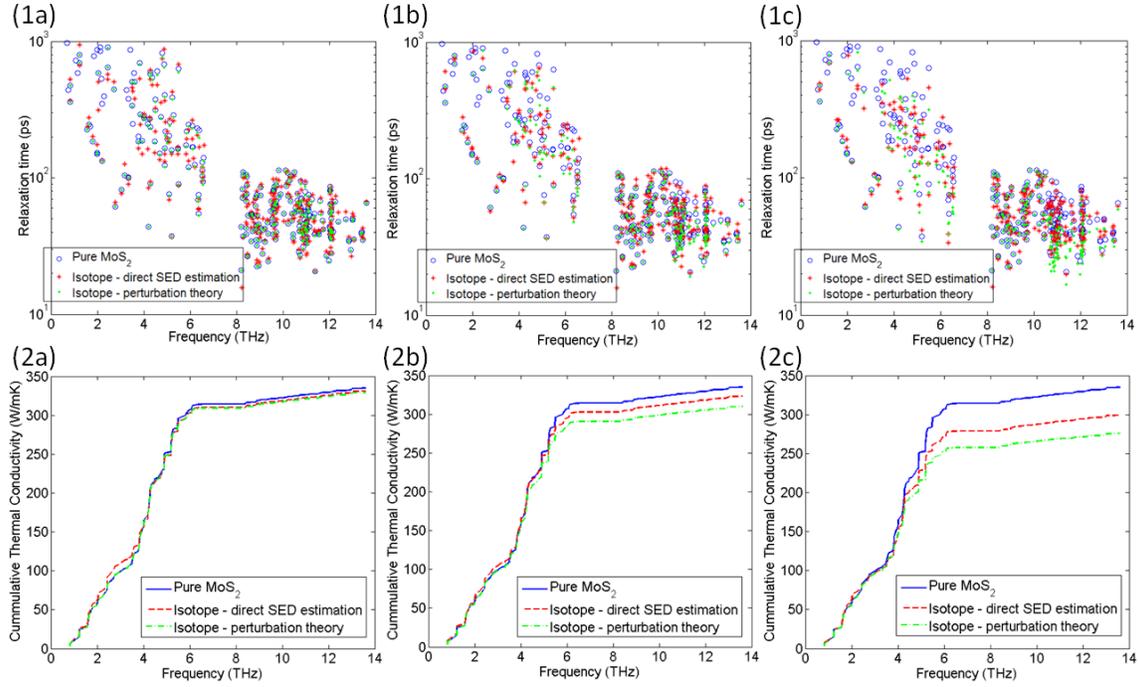

**Figure S6.** Row (1) Relaxation time as a function of frequency, and Row (2) cumulative thermal conductivity as a function of frequency. Letters (a), (b) and (c) represent concentrations of 1%, 5%, and 15 % of Mo isotopes ($Mo^{98}$ in $Mo^{96}$).

### 6. Thermal conductivity from different phonon branches

Here we decompose the total thermal conductivity into the contributions of different phonon branches. Figure S7 shows the thermal conductivity of different phonon branches as a function of length. Unlike graphene that ZA mode dominates the thermal transport, for $MoS_2$, ZA and TA modes here contribute similarly, followed by LA modes. The optical modes contribute the least. It is seen that the optical modes are less influenced by the size effect due to their small relaxation times, group velocities and thus small mean free paths. On the contrary, acoustic modes, which have longer mean free paths, contribute more when size increases.

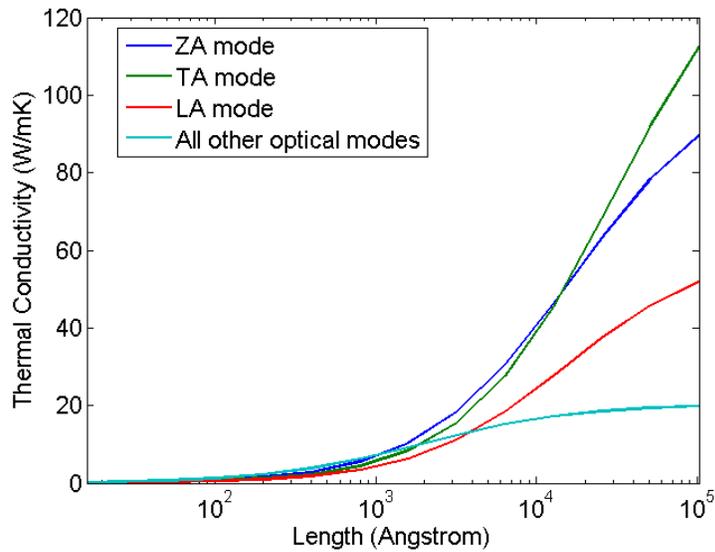

**Figure S7. The thermal conductivity from different phonon branches as a function of length.**